\documentclass[12pt]{article}
\usepackage{amsbsy,graphicx}

\def\ga{\gamma}
\def\de{\delta}
\def\ep{\epsilon}

\def\th{\theta}

\def\Ga{\Gamma}

\def\half{{1\over{2}}}
\def\pd{\partial}
\def\del{{\bf\nabla}}

\newcommand{\beq}{\begin{equation}}
\newcommand{\eeq}{\end{equation}}
\newcommand{\bea}{\begin{eqnarray*}}
\newcommand{\eea}{\end{eqnarray*}}
\newcommand{\beaq}{\begin{eqnarray}}
\newcommand{\eeaq}{\end{eqnarray}}
\newcommand{\vph}{\boldsymbol{\varphi}}
\newcommand{\vrh}{\boldsymbol{\rho}}
\newcommand{\val}{{\boldsymbol{\alpha}}}

\newcommand{\vaa}{\boldsymbol{a}}
\newcommand{\vom}{\boldsymbol{\omega}}

\newcommand{\vP}{{\bf P}}
\newcommand{\vQ}{{\bf Q}}
\newcommand{\bfe}{{\bf e}}

\newcommand{\bds}[1]{\boldsymbol{#1}}

\newcommand{\T}{\mathcal{T}}
\newcommand{\mbar}{{\overline m}}

\begin{document}
\begin{flushright}APCTP-2000-003\\KIAS-P00011\\LPM 00-04\end{flushright}
\centerline{\Large \bf Reflection Amplitudes in Non-Simply Laced Toda Theories}
\centerline{\Large \bf and Thermodynamic Bethe Ansatz}
\vskip 1cm
\centerline{\large Changrim Ahn$^{1,2}$, P.\ Baseilhac$^3$, V. A. Fateev$^{4}$,
Chanju Kim$^{5}$, and Chaiho Rim$^{6,2}$}
\vskip 1cm
\centerline{\it$^{1}$Department of Physics, Ewha Womans University}
\centerline{\it Seoul 120-750, Korea}
\vskip .4cm
\centerline{\it$^{2}$Asia Pacific Center for Theoretical Physics}
\centerline{\it Yoksam-dong 678-39, Seoul, 135-080, Korea}
\vskip .4cm
\centerline{\it$^{3}$Department of Mathematics, University of York}
\centerline{\it Heslington, York YO105DD, United Kingdom}
\vskip .4cm
\centerline{\it$^{4}$Laboratoire de Physique Math\'ematique,
Universit\'e Montpellier II}
\centerline{\it Place E. Bataillon, 34095 Montpellier, France}
\centerline{\it and}
\centerline{\it Landau Institute for Theoretical Physics}
\centerline{\it Kosygina 2, 117334 Moscow, Russia}
\vskip .4cm
\centerline{\it $^{5}$ School of Physics, Korea Institute for Advanced Study}
\centerline{\it Seoul, 130-012, Korea}
\vskip .4cm
\centerline{\it $^{6}$ Department of Physics, Chonbuk National University}
\centerline{\it Chonju 561-756, Korea}
\vskip 0.7cm
\centerline{\small PACS: 11.25.Hf, 11.55.Ds}
\vskip 0.7cm
\centerline{\bf Abstract}
We study the ultraviolet asymptotics in non-simply laced affine Toda 
theories considering them as perturbed non-affine Toda theories,
which possess the extended conformal symmetry.
We calculate the reflection amplitudes, in non-affine Toda theories
and use them to derive the quantization condition for the 
vacuum wave function, describing zero-mode dynamics.
The solution of this quantization conditions for the ground state energy  
determines the UV asymptotics of the effective central charge.
These asymptotics are in a good agreement with Thermodynamic
Bethe Ansatz(TBA) results.
To make the comparison with TBA possible, we give the exact relations 
between parameters of the action  and masses of particles
as well as the bulk free energies for non-simply laced affine Toda
theories.

\newpage
{\small\section{Introduction}}
There is a large class of massive 2D integrable quantum field theories (IQFTs),
which can be considered as perturbed conformal field theories (CFTs) 
\cite{sasha}. The ultraviolet (UV) behavior of these  IQFTs is encoded in the
 CFT data while their long distance properties are defined by the S-matrix
data. If the basic CFT admits the representation of the primary fields
of full symmetry algebra in terms of the exponential fields
the CFT data include ``reflection amplitudes". These functions define the
linear transformations between different exponential fields, corresponding
to the same primary field. Reflection amplitudes  play the crucial role
for the calculation of the one point functions \cite{FLZZ} as well as for 
the description of the zero mode dynamics \cite{ZamZam, AKR, AFKRY} in  
integrable
perturbed CFTs. In particular, the zero mode dynamics determines the UV 
asymptotics of the ground state energy $E(R)$ (or effective central charge
$c_{\rm eff}(R)$) for the system on the circle of size $R$. The function
$c_{\rm eff}(R)$ admits in this case the UV series expansion in the inverse
powers of $\log(1/R)$. The solution of the quantization condition for the
vacuum wave function (which can be written in terms of the reflection 
amplitudes), supplemented with the exact relations between the parameters
of the action and the masses of the particles determines 
all logarithmic terms in this UV expansion.

The effective central charge $c_{\rm eff}(R)$ in IQFT can be calculated 
independently from the S-matrix data using the TBA method \cite{yang,alyosha}.
At small $R$ its asymptotics can be compared with that following from the CFT
data. In the case when the basic CFT is known the agreement of both approaches
can be considered as nontrivial test for the S-matrix amplitudes in IQFT.
The corresponding analysis based on the both approaches was previously done
for the sinh-Gordon \cite{ZamZam}, super-symmetric sinh-Gorgon,
Bullough-Dodd \cite{AKR} models and simply-laced affine Toda
field theories (ATFTs) \cite{AFKRY}.

In this paper we study the UV behavior of the effective central charge 
in ATFTs associated with non-simply laced Lie algebras. These IQFTs have 
two different classical limits. Namely, the weak and strong coupling
limits correspond to the dual pairs of affine Toda theories. As a result,
the mass ratios in these IQFTs depend on the coupling constant and flow
from the classical values characteristic for Lie algebra  $G$%
\footnote{Throughout the paper, we denote an untwisted algebra as $G$,
while $G^\vee$ refers to a twisted one.} to the same values for the dual
algebra $G^\vee$ \cite{DGZ}. The number of particles in ATFTs is equal to
the rank $r$ of $G$. For large $r$ the numerical analysis of TBA equations,
especially in the UV region, becomes rather complicated. 
The analytical approach to the TBA equations \cite{martins, FKS} 
does not give, at present, the regular UV expansion. So, it is useful to 
have the full logarithmic expansion for $c_{\rm eff}(R)$ following from
CFT data. The agreement of this expansion with the TBA results confirms the 
$S$-matrix as well as the relations between the parameters of the action
and masses of particles in non-simply laced ATFT.

The remarkable feature of ATFT is that effective central charge calculated from the CFT data with subtracted bulk free energy term (like in TBA approach) gives
a good agreement with the TBA results even outside the UV region (at
$R \sim {\cal O}(1)$). This ``experimental" fact still needs the explanation.

The rest of the paper is organized as follows. After introduction of some basic 
notations we give the exact relations between the parameters of the action
and masses of particles in non-simply laced ATFTs. Then following the 
procedure of ref. \cite{AFKRY}, we obtain the reflection amplitudes and
quantization conditions for the wave function, describing the vacuum zero mode
dynamics. Using these results we calculate the UV asymptotics of the effective
central charges for ATFTs and compare these asymptotics with numerical data
following from TBA equations.  We omit here the details, which can be found
in ref. \cite{AFKRY}, devoted to the analysis of UV asymptotics in 
simply laced ATFTs.

{\small\section{Mass-$\boldsymbol{\mu}$ Relations and Reflection Amplitudes}}

The ATFTs corresponding to Lie algebra $G$ is described by the action
\beq
{\cal A}=\int d^2x\left[{1\over{8\pi}}(\pd_{\mu}\vph)^2
+\sum_{i=1}^{r}\mu_i e^{b\bfe_i\cdot\vph}+\mu_0 e^{b\bfe_0\cdot\vph}
\right],\label{action}
\eeq
where $\bfe_i,\ i=1,\ldots,r$ are the simple roots of the Lie algebra
$G$ of rank $r$ and $-\bfe_0$ is a maximal root, satisfying the relation:
\beq
\sum_{i=0}^{r}n_i\bfe_i=0,\qquad n_0=1.
\label{maxroot}
\eeq
Non-simply laced ATFTs have standard simple roots with $\bfe_i^2=2$
and nonstandard simple roots with $\bfe_i^2 \equiv \xi^2 (\neq 2)$. We 
choose the corresponding parameters $\mu_i$ as $\mu$ (for standard roots) 
and $\mu'$ (for nonstandard ones) respectively%
\footnote{We choose the convention that the length squared of the long 
roots are four for $C_r^{(1)}$ and two for the other untwisted algebras.}.

In the case of non-simply laced ATFTs, the exact mass ratios are 
different from the classical ones and get quantum corrections \cite{DGZ, CDS}.
To describe the spectrum it is convenient to introduce the notations:
\beq \label{largeh}
 B = \frac{b^2}{1+b^2}, \qquad H = \frac{h + b^2 h^\vee}{1+b^2},
\eeq
where $h$ and $h^\vee$ are Coxeter and dual Coxeter numbers of the algebra.
Then the spectrum of ATFTs can be expressed in terms of one mass parameter
$\mbar$ as:
\beaq
B_r^{(1)}: &&  M_r=\mbar, \qquad  M_a =2\mbar\sin(\pi a/H),
                                \qquad a=1,2,\ldots,r-1 \nonumber\\
C_r^{(1)}: &&  M_a =2\mbar\sin(\pi a/H), \qquad a = 1,2,\ldots,r \nonumber\\
G_2^{(1)}: &&  M_1=\mbar, \qquad M_2 =2\mbar\cos(\pi(1/3 -1/H))\nonumber\\
F_4^{(1)}: &&  M_1=\mbar,\qquad M_2 =2\mbar\cos(\pi(1/3 -1/H)),\nonumber\\
           &&  M_3 =2\mbar \cos(\pi(1/6 -1/H)),\qquad
               M_4 = 2M_2 \cos(\pi/H)\,.   
\eeaq  
The relation between the parameter $\mbar$ in the above spectra and
the parameters $\mu_i$  in the action (\ref{action}) can be 
obtained by Bethe Ansatz method (see for example \cite{mumass,fateev}). 
The corresponding analysis gives:
\beq \label{mmu}
\prod_{i=0}^{r}[-\pi\mu_i\gamma(1+\bfe_i^2b^2/2)]^{n_i}
   =\left[\frac{\mbar k(G)}{2} \Gamma\left(\frac{1-B}{H}\right) 
      \Gamma\left(1+\frac{B}{H}\right)\right]^{2H(1+b^2)} \,,
\eeq
where, as usual $\ga(x)=\Ga(x)/\Ga(1-x)$,
and $k(G)$ is a function depending on the algebra:
\beaq \label{kofg}
k(B_r^{(1)}) = \frac{2^{-2/H}}{\Gamma(1/H)}, \qquad &&
k(C_r^{(1)}) = \frac{2^{2B/H}}{\Gamma(1/H)}\,, \nonumber\\
k(G_2^{(1)}) = \frac{\Gamma(2/3)}{2 \Gamma(1/2)\Gamma(1/6+1/H)}, \qquad &&
k(F_4^{(1)}) = \frac{\Gamma(2/3)}{2 \Gamma(1/2)\Gamma(1/6+1/H)}\,.
\eeaq
The similar relations for the dual ATFTs can be easily obtained 
from Eqs.(\ref{mmu}, \ref{kofg}) if we use the duality relations 
for the parameters $\mu_i$ and $\mu_i^\vee$ corresponding to 
the dual pairs of ATFTs:
\beq \label{dur}
\pi\mu_i\gamma\left(\frac{\bfe_i^2b^2}{2}\right) =
\left(\pi\mu_i^\vee\gamma\left(\frac{2}{\bfe_i^2b^2}\right)
\right)^{\bfe_i^2b^2/2}
\eeq

The ATFTs can be considered as perturbed CFTs. 
Without the last term with the zeroth root $\bfe_0$, the action
in Eq.(\ref{action}) describes the non-affine Toda theory (NATT),
which is conformal.
To describe the generator of conformal symmetry we introduce
the complex coordinates $z=x_1+i x_2$ and ${\overline z}=x_1-i x_2$
and vector:
\beq 
\vQ=b\vrh + \frac{1}{b}\vrh^\vee, \qquad 
\vrh=\half\sum_{\val>0}\val, \qquad 
\vrh^\vee=\half\sum_{\val>0}\val^\vee,
\eeq
where the sum in definition of Weyl vector $\vrh$ ($\vrh^\vee$) runs over
all positive roots $\val$ (co-roots $\val^\vee$) of $G$.

The holomorphic stress-energy tensor
\beq
T(z)=-\half(\pd_z\vph)^2+\vQ\cdot\pd_z^2\vph
\eeq
ensures the local conformal invariance of the NATT with the central
charge $c=r+12\vQ^2$.

Besides the conformal invariance the NATT possesses extended
symmetry generated by $W(G)$-algebra.
The full chiral $W(G)$-algebra contains $r$ holomorphic fields
$W_j(z)$ ($W_2(z)=T(z)$) with spins $j$ which follows the
exponents of Lie algebra $G$.  
The primary fields $\Phi_w$ of $W(G)$ algebra are classified
by  $r$ eigenvalues $w_j,\ j=1,\ldots,r$ of the operator
$W_{j,0}$ (the zeroth Fourier component of the current  $W_j(z)$):
\beq
W_{j,0}\Phi_w=w_j\Phi_w,\qquad
W_{j,n}\Phi_w=0,\quad n>0.
\eeq
The exponential fields
\beq
V_{\vaa}(x)=e^{(\vQ+\vaa)\cdot\vph(x)}
\label{vertex}
\eeq
are spinless conformal primary fields with dimensions 
$\Delta(\vaa)=w_2(\vaa)=(\vQ^2-\vaa^2)/2$.
The fields $V_{\vaa}$ are also primary 
with respect to all chiral algebra $W(G)$ with the eigenvalues
$w_j$ depending on $\vaa$. 
The functions $w_j(\vaa)$, which define the representation of
$W(G)$-algebra possess the symmetry with respect to the Weyl
group ${\cal W}$ of Lie algebra $G$ \cite{FL, BS}, i.e.
$w_j({\hat s}\vaa)=w_j(\vaa)$; for any 
${\hat s}\in {\cal W}$.
It means that the fields $V_{{\hat s}\vaa}$ for different
${\hat s}\in {\cal W}$ are reflection images of each other
and are related by the linear transformation:
\beq
V_{\vaa}(x)=R_{\hat s}(\vaa)V_{{\hat s}\vaa}(x)
\label{reflection}
\eeq
where $R_{\hat s}(\vaa)$ is the ``reflection amplitude''.

This function plays an important role in the analysis of perturbed CFTs.
It can be calculated by the CFT methods (exactly in the same way as it was done for the simply laced NATTs in \cite{AFKRY}) and has the form:  
\beq
R_{\hat s}(\vaa)=
{A_{{\hat s}\vaa}\over{A_{\vaa}}}
\label{rhats}
\eeq
where 
\beq
A_{\vaa}=\prod_{i=1}^{r}[\pi\mu_i\gamma(\bfe_i^2b^2/2)]^{\vom^\vee_i\cdot\vaa/b}
\prod_{\val>0}\Ga(1-a_{\val^\vee}/b)\Ga(1-a_{\val}b),
\label{aa}
\eeq
here $a_{\val}=\vaa\cdot\val$, $a_{\val^\vee}=\vaa\cdot\val^\vee$ and vectors
$\vom^\vee_i$ are the co-weights of $G$, satisfying the
condition $\vom^\vee_i\cdot\bfe_j=\de_{ij}$

In following we will be interested in the values of function $A(\vP)=A_{i\vP}$ .
We note that in the semiclassical limit ($b\to 0$ with $\vP/b$ fixed) the
functions $A({\hat s}\vP)$ coincide with the amplitudes describing the 
asymptotics of the wave function of quantum mechanical non-affine Toda chain
(\ref{vftc}) (see for example \cite{OlsPer}).

{\small\section{Quantization Condition and UV expansion}}

 Function $A(\vP)$ plays an important role in study of quantum mechanical
problem for zero modes
\beq
\vph_0 = \int_0^{2\pi}\vph(x) \frac{dx_1}{2\pi},
\eeq
of the fields $\vph(x)$ defined
on an infinite cylinder of circumference
$2\pi$ with coordinate $x_2$ along the cylinder playing the role
of imaginary time. In the semiclassical limit $b\to 0$,
where one can neglect the oscillator modes of $\vph(x)$, the Schr\"odinger
equation governing the zero-mode dynamics is given by:
\beq
\left[-{r\over{12}}-\del_{\vph_0}^2
+\sum_{i=1}^{r} 2\pi \mu_i e^{b\bfe_i\cdot\vph_0}\right]
\Psi_{\vP}(\vph_0) = E_0\Psi_{\vP}(\vph_0)
\label{vftc}
\eeq
with the energy
\beq
E_0=-{r\over{12}}+\vP^2.
\label{energy}
\eeq
where the momentum $\vP$ is a real vector.
The full quantum effect can be implemented simply by introducing
the exact reflection amplitudes which take into account also non-zero-mode
contributions \cite{ZamZam}.

The wave function $\Psi_\vP(\vph_0)$ in the asymptotic region (Weyl chamber)
can be found by
using the same arguments as was given in \cite{AFKRY} for simply 
laced NATTs. The only
difference is that there are now two kinds of roots with different lengths. 
Namely, each exponential term $\mu_i e^{b \bfe_i \vph_0}$ in the 
Hamiltonian can be considered as a potential wall normal to the $\bfe_i$ 
direction.  An incident wave is reflected by this wall to the wave with the 
Weyl-reflected momentum. The phase change corresponding to this process should 
be the same as in Liouville field theory. By considering 
the reflections from all  
potential walls, we find that the wave function $\Psi_\vP(\vph_0)$ can be 
written as a superposition of plane waves with the momenta forming the orbit 
of the Weyl group ${\cal W}$ of Lie algebra $G$,
\beq \label{wavefunction}
\Psi_\vP(\vph_0)\simeq\sum_{\hat{s}\in {\cal W}}
A(\hat{s}\vP)e^{i\hat{s}\vP\cdot\vph_0},
\eeq
where 
\beq \label{amplitude}
A(\vP) = \prod_{i=1}^{r}[\pi\mu_i\gamma(\bfe_i^2b^2/2)]^{i\vom^\vee_i\cdot\vP/b}
          \prod_{\val>0}\Gamma(1-iP_{\val}b) \Gamma(1-iP_{\val^\vee}/b),
\eeq
For the Weyl element ${\hat s}_i$, associated with the
simple root $\bfe_i$, the ratio $A(\hat{s}_i\vP)/A(\vP)$ should be given by the
reflection amplitude $S_L(\bfe_i, \vP)$ \cite{ZamZam} of the Liouville field 
theory
\beaq \label{liouville}
\frac{A(\hat{s}_i\vP)}{A(\vP)} &=& S_L(\bfe_i, \vP) \nonumber \\
  &=& [\pi\mu_i\gamma(\bfe_i^2b^2/2)]^{-i\vP\cdot\bfe_i^\vee/b}
       \frac{\Gamma(1+i\vP\cdot\bfe_ib)\Gamma(1+i\vP\cdot\bfe_i^\vee/b)}%
            {\Gamma(1-i\vP\cdot\bfe_ib)\Gamma(1-i\vP\cdot\bfe_i^\vee/b)}.
\eeaq
One can easily check that function $A(\vP)$ satisfies this 
functional equation.
With this function one can proceed to obtain the
scaling functions in the UV region of the ATFTs defined 
on a cylinder with circumference $R\rightarrow 0$. 
The additional  term in the ATFT Lagrangian corresponding to 
the zeroth root $\bfe_0$ introduces new potential wall in that direction.
With this term the Weyl chamber is now closed and 
the momentum $\vP$ of the wave 
function should be quantized. It depends on the size of the enclosed region,
which is proportional to $\log(1/R)$.
This quantized momentum $\vP(R)$ defines the scaling 
function $c_{\rm eff}$ in the 
UV region by Eq.(\ref{energy}).

It is convenient to rescale back the size of the system from $R$ to $2\pi$. 
This leads to the following rescaling of the parameters $\mu_i$ in the 
action (\ref{action}):
\beq \label{nu}   
\mu_i \rightarrow \nu_i= \mu_i \left(\frac{R}{2\pi}\right)^{2+b^2\bfe_i^2}\,,
\eeq
In the UV limit the size of enclosed region is rather big and we can neglect 
the subtleties of interaction (which give only exponential corrections)
taking into account only the phase shifts coming from the reflections of 
the waves by the potential walls. 
Since the additional potential term is not different from the others, the
amplitude $A(\hat{s}\vP)$  
with the momentum $\hat{s}\vP$ (where $\hat{s}$ is
an arbitrary element of Weyl group) has to satisfy also the reflection relation 
(\ref{liouville})  with respect to the zeroth root $\bfe_0$
\beq \label{ratios}
\frac{A(\hat{s}_0\hat{s}\vP)}{A(\hat{s}\vP)} = S_L(\bfe_0,\hat{s}\vP)\,.
\eeq
Inserting Eqs.(\ref{amplitude}) and (\ref{liouville}) into
Eq.(\ref{ratios}), we obtain the condition for $\vP$. After some
transformations (see ref.\cite{AFKRY} for details), it can be written
in the form:
\beq \label{quant2}
\left[\prod_{i=0}^{r}\left(\pi\nu_i\gamma(\bfe_i^2b^2/2)\right)^{n_i}  
\right]^{i\vP\cdot\hat{s}\bfe_0^\vee/b}
\prod_{\val>0}\left[
\frac{{\cal G}(\val,\vP)}{{\cal G}(\val,-\vP)}
\right]^{\val\cdot\hat{s}\bfe_0^\vee} = 1\,,
\eeq
where $\nu_i$ are defined by Eq.(\ref{nu}) and
\[
{\cal G}(\val,\vP) = \Gamma(1 - i P_\val b) \Gamma(1 - i P_{\val^\vee} /b)\,.
\]
For the lowest energy state, Eq.(\ref{quant2}) reduces to the following
equation:
\beq \label{quantization}
L\vP = 2\pi\vrh - \sum_{\val>0}\val\de(\val, \vP)\,,
\eeq
where
\beq \label{L}
L=-\frac2b(h + b^2 h^\vee)\ln\frac{R}{2\pi} 
  -\frac1b
 \ln\left[\prod_{i=0}^{r}\left(\pi\mu_i\gamma(\bfe_i^2b^2/2)\right)^{n_i}
\right],
\eeq
and
\beq \label{delta}
\de(\val,\vP)=-i \log\frac{\Gamma(1+i P_\val b)\Gamma(1+i P_{\val^\vee}/b)}%
                  {\Gamma(1-i P_\val b)\Gamma(1-i P_{\val^\vee}/b)}\,.
\eeq
This is the quantization condition for the momentum $\vP$ 
in the UV region $R\rightarrow 0$. 
The ground state energy of the system on the circle of size $R$ is 
then given by
\beq
E(R)=-{\pi c_{\rm eff}\over{6R}}\quad
{\rm with}\quad
c_{\rm eff}=r-12\vP^2
\label{grenergy}
\eeq
where $\vP$ is the solution of Eq.(\ref{quantization}).

In the UV region we can solve Eq.(\ref{quantization}) perturbatively by
expanding $\de(\val, \vP)$ in powers of $P_\val$,
\beq \label{expand}
\de(\val, \vP) = \de_1(\val,b)P_\val 
                + \de_3(\val,b)P_\val^3 + \de_5(\val,b) P_\val^5 \cdots\,,
\eeq
where the coefficients $\de_1(\val,b)$ and $\de_s(\val,b)$, $s=3,5$ are: 
\beq \label{deltas}
\de_1(\val,b) = 
      -2\gamma_E \left(b + \frac{2}{\val^2b}\right),\qquad
\de_s(\val,b) =(-)^{\frac{s-3}{2}}\cdot\frac{2}{s}\zeta(s)\left(b^s 
           + \left(\frac{2}{\val^2b}\right)^s\right). 
\eeq
Using the relations: 
$
\sum_{\val>0} (\val)^a(\val)^b = h^\vee\de^{ab},
$
and
$
\sum_{\val>0} (\val)^a(\val^\vee)^b = h\de^{ab},
$
we obtain that:
\[
l\vP = 2\pi\vrh-\sum_{\val>0}\de_3(\val,b)\val P_{\val}^3
       -\sum_{\val>0}\de_5(\val,b)\val P_{\val}^5 - \cdots\,,
\]
with
\beq \label{smalll}
l= L - 2 \gamma_E (bh^\vee + h/b)\equiv L-L_0\,.
\eeq
The above equation can be solved iteratively in powers of $1/l$. 
Inserting the solution into Eq.(\ref{grenergy}), we find:
\beaq \label{ceff}
c_{\rm eff}&=&r-r(h+1){h^\vee}\left(\frac{2\pi}{l}\right)^2
+\frac{8}{\pi}\zeta(3)
   [C_4(G^\vee)b^3 + C_4(G)/b^3]\left(\frac{2\pi}{l}\right)^5 \nonumber \\
&&-\frac{24}{5\pi}\zeta(5)
   [C_6(G^\vee)b^5 + C_6(G)/b^5]\left(\frac{2\pi}{l}\right)^7
+{\cal O}(l^{-8})\,,
\eeaq
where the coefficients $C(G)$ are defined as:
\bea
&&C_4(G) = \sum_{\val > 0} \rho_\val \rho_{\val^\vee}^3, \qquad
C_4(G^\vee) = \sum_{\val > 0} \rho_{\val}^4, \nonumber\\
&&C_6(G) = \sum_{\val > 0} \rho_\val \rho_{\val^\vee}^5, \qquad
C_6(G^\vee) = \sum_{\val > 0} \rho_{\val}^6.
\eea
For simply laced algebras, these coefficients were calculated in  
\cite{AFKRY} and have the values:
\beaq
C_4(A_{n-1}^{(1)}) &=& \frac{1}{60} n^2 (n^2 -1)(2n^2-3)\,, \nonumber \\
C_6(A_{n-1}^{(1)}) &=& \frac{1}{168}n^2 (n^2 -1)(n^2-2)(3n^2-5)\,,\nonumber\\
C_4(D_n^{(1)}) &=& \frac{1}{30}(16n^3-45n^2+27n+8)n(n-1)(2n-1)\,,\nonumber\\
C_6(D_n^{(1)}) &=& \frac{1}{42}(48n^5-213n^4+262n^3+6n^2-101n-32)n(n-1)(2n-1).
\eeaq
For the non-simply laced algebras $B_n^{(1)}$ and $C_n^{(1)}$, we can express
the results through these values. Namely, we find:
\beaq
&C_i(B_n^{(1)}) = \half C_i(A_{2n-1}^{(1)}),\qquad &
  C_i(B_n^{(1)\, \vee}) = C_i(D_{n+1/2}^{(1)}), \nonumber\\
&C_i(C_n^{(1)}) = C_i(D_{n+1}^{(1)}), \qquad &
  C_i(C_n^{(1)\, \vee}) = C_i(D_{-n}^{(1)}), \qquad (i=4,6).
\eeaq
For exceptional algebras $G_2^{(1)}$ and $F_4^{(1)}$, we obtain:
\beaq
&C_4(G_2^{(1)}) = \frac{1}{3}C_4(D_4^{(1)}) = 392,
\qquad & C_4(G_2^{(1)\,\vee})=\frac{980}{9},\nonumber\\
&C_6(G_2^{(1)}) = \frac{1}{3}C_6(D_4^{(1)}) =7386, \qquad 
       & C_6(G_2^{(1)\,\vee}) = \frac{199516}{243}, \nonumber\\
&C_4(F_4^{(1)}) = \frac{1}{2}C_4(E_6^{(1)}) = 27378,\qquad 
       & C_4(F_4^{(1)\,\vee})=\frac{22815}{2},\nonumber\\
&C_6(F_4^{(1)}) = \frac{1}{2}C_6(E_6^{(1)})=2203578, \qquad 
       & C_6(F_4^{(1)\,\vee}) = \frac{4052763}{8}.
\eeaq
We note that above equations relating coefficients $C_i(G)$ for different
Lie algebras follow from the similar exact relations between the ground state 
energies $e(G)$ of quantum affine Toda chains associated with these Lie 
algebras. These exact relations are valid if the parameters $\mu$, $\mu'$ for
non-simply laced Lie algebras and corresponding parameter $\mu_{sl}$ for simply 
laced ones satisfy the condition: $\mu^{h-z}(2\mu'/\xi^2)^z =\mu_{sl}^h$,
where $z = \frac{2(h - h^\vee)}{2 - \xi^2}$.

{\small\section{Comparison with TBA results}}
The effective central charge calculated above from the CFT data (reflection 
amplitudes) can be compared with the same function determined from
numerical solution of the TBA equations for ATFTs. Namely:
\beq  \label{ceff_tba}
c_{\rm eff}^{\rm(TBA)}(R) = \sum_{i=1}^r\frac{3Rm_i}{\pi^2}
\int \cosh\theta \log\left(1+ e^{-\ep_i(\th,R)}\right)d\theta.
\eeq
where functions $\ep_i(\theta,R)$ ($i=1, \cdots, r$) satisfy the system of 
$r$ coupled integral equations:
\beq
m_i R \cosh\theta=\ep_i (\theta,R)+
\sum_{j=1}^r\int\varphi_{ij} (\theta-\theta') 
\log\left(1+ e^{-\ep_i(\th',R)}\right){d\theta'\over 2\pi},
\label{toda_tba}
\eeq
with the kernels $\varphi_{ij}$, equal to the logarithmic 
derivatives of the $S$-matrices $S_{ij} (\th)$ of ATFTs, conjectured
in \cite{DGZ, CDS}.

The function $E^{(TBA)}(R)$ defined from the TBA equations differs
from the ground state energy $E(R)$ of the system on the circle of size $R$
by the bulk term: $E^{(TBA)}(R)=E(R)-fR$, where $f$ is a specific
bulk free energy \cite{alyosha}. To compare the same functions we
should subtract this term from the function $E(R)$ defined by Eq.(\ref{ceff}) 
i.e.   
\beq \label{deltac}
  c_{\rm eff}^{\rm (TBA)}(R)=c_{\rm eff}^{\rm (RA)}(R)  
                                   + \frac{6R^2}{\pi}f(G)\,.
\eeq
The specific bulk free energy $f(G)$ for non-simply laced ATFTs
can be calculated by Bethe Ansatz method with the result:
\beaq \label{fg}
f(G)&=&\frac{\mbar^2 \sin(\pi/H)}{8\sin(\pi B/H)\sin(\pi(1-B)/H)}\,,\qquad 
     G=B_r^{(1)},\ C_r^{(1)},\nonumber\\
f(G)&=&\frac{\mbar^2 \cos(\pi(1/3-1/H))}{16\cos(\pi/6)\sin(\pi B/H)
                                           \sin(\pi(1-B)/H)}\,,\qquad 
     G=G_2^{(1)},\ F_4^{(1)}.
\eeaq
The contribution of bulk term $f(G)$  becomes quite essential 
at $R\sim {\cal O}(1)$.

The TBA Eqs.(\ref{toda_tba}) were solved numerically for 
non-simply laced algebras, $B_2^{(1)}$ 
($=C_2^{(1)}$), $B_3^{(1)}$, $B_4^{(1)}$, $C_3^{(1)}$, 
$C_4^{(1)}$, $G_2^{(1)}$ and $F_4^{(1)}$.  
The effective central charge $c_{\rm eff}^{\rm (TBA)}(R)$ was then computed 
from Eq.(\ref{ceff_tba}) for many different values of parameter $\mbar R$.
After taking into account the bulk term, 
the numerical solution for $c_{\rm eff}^{\rm (TBA)}(R)$ was fitted
with the expansion (\ref{ceff}) (neglecting higher order terms in $1/l$):
\beq \label{expan}
c_{\rm eff}^{\rm (RA)}(R) = 
    r - r(h+1){h^\vee}\left(\frac{2\pi}{l}\right)^2
   +c_5\left(\frac{2\pi}{l}\right)^5 + c_7\left(\frac{2\pi}{l}\right)^7\,.
\eeq
with fitting parameters $L_0$, $c_5$ and $c_7$, where parameter $L_0$ is
defined by the Eq.(\ref{smalll}). The exact values of these parameters can be 
easily identified from Eqs.(\ref{smalll}) and (\ref{ceff}). To compare the
expansion (\ref{expan}) with TBA results we use the relations (\ref{mmu}) 
between parameters $\mu_i$ in the action and the parameter $\mbar$ 
characterizing  the spectrum of particles. It gives the following expression
for function $L(R)$ in Eqs.(\ref{smalll}):
\beq
L=-\frac2b(h + b^2 h^\vee)
     \ln\left[\frac{\mbar R}{4\pi} k(G) 
              \Gamma\left(\frac{1-B}{H}\right) 
              \Gamma\left(1+\frac{B}{H}\right)\right]
  +\frac{2}{b}\ln(b^{2h}(\xi^2/2)^{z})\,.
\eeq

Tables 1--3 show the values of parameters $L_0$, $c_5$ and $c_7$ obtained
numerically from TBA equations (denoted with the superscript (TBA)) and 
those obtained analytically (Eqs.(\ref{smalll}) and (\ref{ceff})) from 
reflection amplitudes (denoted with the superscript (RA))  
for $C_2^{(1)}$, $C_3^{(1)}$, 
$C_4^{(1)}$, $B_3^{(1)}$, $B_4^{(1)}$, $G_2^{(1)}$ and $F_4^{(1)}$ ATFTs
 with different values of the parameter $B$. 
We see that both data are in excellent agreement. 
(Relatively poor accuracy for $c_7$ is mainly due to the limitation of 
numerical accuracy and the influence 
of higher order term (${\cal O}(l^{-8})$) in the expansion (\ref{expan}).)
This agreement supports the approach based on the reflection amplitudes,
$\mu$-${\overline m}$ relations and quantization conditions as well as
the $S$-matrices for non-simply laced ATFTs.
\begin{table}
\begin{tabular}{||c|c|c|c|c|c|c||} \hline
\rule[-.2cm]{0cm}{0.8cm}B&
     0.3 &     0.4 &     0.5 &     0.6 &     0.7 & 0.8 \\ \hline
$L_0^{\rm (RA)}(C_2^{(1)})$ &
 11.5882 & 11.3111 & 11.5443 & 12.2537 & 13.6035 & 16.162 \\
$L_0^{\rm (TBA)}(C_2^{(1)})$ & 
 11.5882 & 11.3111 & 11.5443 & 12.2537 & 13.6035 & 16.162 \\ \hline
$L_0^{\rm (RA)}(C_3^{(1)})$ & 
 16.6266 & 16.0240 & 16.1620 & 16.9666 & 18.6419 & 21.9342 \\
$L_0^{\rm (TBA)}(C_3^{(1)})$ & 
 16.6266 & 16.0240 & 16.1620 & 16.9666 & 18.6419 & 21.9319 \\ \hline
$L_0^{\rm (RA)}(C_4^{(1)})$ & 
 21.6649 & 20.7370 & 20.7798 & 21.6796 & 23.6803 & 27.7064 \\
$L_0^{\rm (TBA)}(C_4^{(1)})$ & 
 21.6649 & 20.7370 & 20.7798 & 21.6795 & 23.6802 & 27.5 \\ \hline
$L_0^{\rm (RA)}(B_3^{(1)})$ & 
 14.3593 & 13.1962 & 12.6987 & 12.7250 & 13.3516 & 15.0076 \\
$L_0^{\rm (TBA)}(B_3^{(1)})$ & 
 14.3589 & 13.1962 & 12.6987 & 12.7250 & 13.3516 & 15.0076 \\ \hline
$L_0^{\rm (RA)}(B_4^{(1)})$ & 
 19.3977 & 17.9092 & 17.3165 & 17.4379 & 18.3900 & 20.7798 \\
$L_0^{\rm (TBA)}(B_4^{(1)})$ & 
 19.32   & 17.9089 & 17.3165 & 17.4379 & 18.3900 & 20.7792 \\ \hline
$L_0^{\rm (RA)}(G_2^{(1)})$ & 
 13.6035 & 12.2537 & 11.5443 & 11.3111 & 11.5882 & 12.6987 \\
$L_0^{\rm (TBA)}(G_2^{(1)})$ & 
 13.3    & 12.2529 & 11.5443 & 11.3111 & 11.5882 & 12.6987 \\ \hline
$L_0^{\rm (RA)}(F_4^{(1)})$ & 
 27.9629 & 25.4499 & 24.2431 & 24.0360 & 24.9398 & 27.7064 \\
$L_0^{\rm (TBA)}(F_4^{(1)})$ & 
         & 25.1    & 24.238  & 24.0360 & 24.9398 & 27.5 \\ \hline
\end{tabular}
\caption{$L_0^{\rm (RA)}$ vs.
$L_0^{\rm (TBA)}$ for non-simply laced ATFTs.}
\end{table}
\begin{table}
\begin{tabular}{||c|c|c|c|c|c|c||} \hline
\rule[-.2cm]{0cm}{0.8cm}B &
     0.3 &     0.4 &     0.5 &     0.6 &     0.7 & 0.8 \\ \hline
$c_5^{\rm (RA)}(C_2^{(1)})$ &
 1569.60 & 1242.16 & 1438.68 & 2183.98 & 3961.81 & 8713.17 \\
$c_5^{\rm (TBA)}(C_2^{(1)})$ &
 1567.   & 1240.   & 1437.   & 2182.   & 3959.   & 8708.   \\ \hline
$c_5^{\rm (RA)}(C_3^{(1)})$ &
 15018.6 & 10858.7 & 11399.2 & 16288.0 & 28809.0 & 62845.7 \\
$c_5^{\rm (TBA)}(C_3^{(1)})$ &
 15000   & 10840   & 11380   & 16270   & 28790   & 68000   \\ \hline
$c_5^{\rm (RA)}(C_4^{(1)})$ &
 76141.1 & 52782.9 & 52563.7 & 72394.9 & 125955. & 273273. \\
$c_5^{\rm (TBA)}(C_4^{(1)})$ &
 76100   & 52700   & 52500   & 72300   & 125990  &         \\ \hline
$c_5^{\rm (RA)}(B_3^{(1)})$ &
 8260.97 & 4765.97 & 3488.60 & 3541.95 & 5151.97 & 10444.4 \\
$c_5^{\rm (TBA)}(B_3^{(1)})$ &
 8500    & 4761.   & 3484.   & 3538.   & 5147.   & 10439.  \\ \hline
$c_5^{\rm (RA)}(B_4^{(1)})$ &
 48261.8 & 28350.9 & 21550.3 & 22970.0 & 34594.6 & 71159.4 \\
$c_5^{\rm (TBA)}(B_4^{(1)})$ &
         & 29000   & 21530   & 22940   & 34560   & 73000   \\ \hline
$c_5^{\rm (RA)}(G_2^{(1)})$ &
 4370.29 & 2385.82 & 1533.23 & 1265.48 & 1524.65 & 2816.47 \\ 
$c_5^{\rm (TBA)}(G_2^{(1)})$ &
         & 2600    & 1532.   & 1264.   & 1523.   & 2814.   \\ \hline
$c_5^{\rm (RA)}(F_4^{(1)})$ &
 308495. & 172966. & 118723. & 109767. & 147970. & 289824. \\
$c_5^{\rm (TBA)}(F_4^{(1)})$ &
         &         & 160000  & 110055. & 148248. &         \\ \hline
\end{tabular}
\caption{$c_5^{\rm (RA)}$ vs.
$c_5^{\rm (TBA)}$ for non-simply laced ATFTs.}
\end{table}
\begin{table}
\begin{tabular}{||c|c|c|c|c|c|c||} \hline
\rule[-.2cm]{0cm}{0.8cm}B &
     0.3 &     0.4 &     0.5 &     0.6 &     0.7 & 0.8 \\ \hline
$c_7^{\rm (RA)}(C_2^{(1)})$ &
--12262.9 & --6566.01& --9109.78& --21843.4& --64594.2& --247955. \\ 
$c_7^{\rm (TBA)}(C_2^{(1)})$ &
--14000   & --8000   & --10600  & --23800  & --68000  & --256000  \\ \hline
$c_7^{\rm (RA)}(C_3^{(1)})$ &
--304830. & --137513.& --148031.& --324132.& --944077.& --3617480 \\
$c_7^{\rm (TBA)}(C_3^{(1)})$ &
--330000  & --160000 & --170000 & --350000 & --990000 &          \\ \hline
$c_7^{\rm (RA)}(C_4^{(1)})$ &
--2878340 & --1216910& --1148970& --2366240& --6815270& --26079300\\ 
$c_7^{\rm (TBA)}(C_4^{(1)})$ &
--3000000 & --1400000& --1300000& --2600000& --7000000&           \\ \hline
$c_7^{\rm (RA)}(B_3^{(1)})$ &
--174026. & --60330.1& --29061.1& --30298.6& --76073.1& --264814. \\
$c_7^{\rm (TBA)}(B_3^{(1)})$ &
         & --65000  & --33000  & --34000  & --76000  & --270000  \\ \hline
$c_7^{\rm (RA)}(B_4^{(1)})$ &
--1847280 & --650082.& --338009.& --404020.& --1004420& --3769680 \\
$c_7^{\rm (TBA)}(B_4^{(1)})$ &
         &         & --380000 & --450000 & --1100000& --2200000 \\ \hline
$c_7^{\rm (RA)}(G_2^{(1)})$ &
--97474.0 & --32718.2& --13002.5& --7830.99& --12225.2& --41991.4 \\
$c_7^{\rm (TBA)}(G_2^{(1)})$ &
         &         & --14000  & --9000   & --14000  & --44000   \\ \hline
$c_7^{\rm (RA)}(F_4^{(1)})$ &
--2913070 & --9911730& --4293750& --3478610& --7094660& --25792400 \\
$c_7^{\rm (TBA)}(F_4^{(1)})$ &
         &         &         & --3410000& --7090000&         \\ \hline
\end{tabular}
\caption{$c_7^{\rm (RA)}$ vs.
$c_7^{\rm (TBA)}$ for non-simply laced ATFTs.}
\end{table}

In Fig.1, we  plot the functions $c_{\rm eff}^{\rm (TBA)}(R)$ and  
$c_{\rm eff}^{\rm (RA)}(R)$ 
for different ATFTs setting ${\overline m}=1$.  
The first function is computed numerically from TBA equations. 
The second one is calculated using Eqs.(\ref{quantization}) 
and (\ref{grenergy}), based 
on the reflection amplitudes, with taking into account the bulk 
free energy term according to Eq.(\ref{deltac}). For
all models, the two curves are almost identical without essential
difference in the graphs even at $R\sim {\cal O}(1)$. This good
agreement outside the UV region looks not to be accidental. However,
at present, we have no satisfactory explanation of this interesting 
phenomena in ATFTs.
\begin{figure}
\rotatebox{-90}{\resizebox{!}{15cm}{\scalebox{0.1}{%
{\includegraphics[4cm,2cm][22cm,25cm]{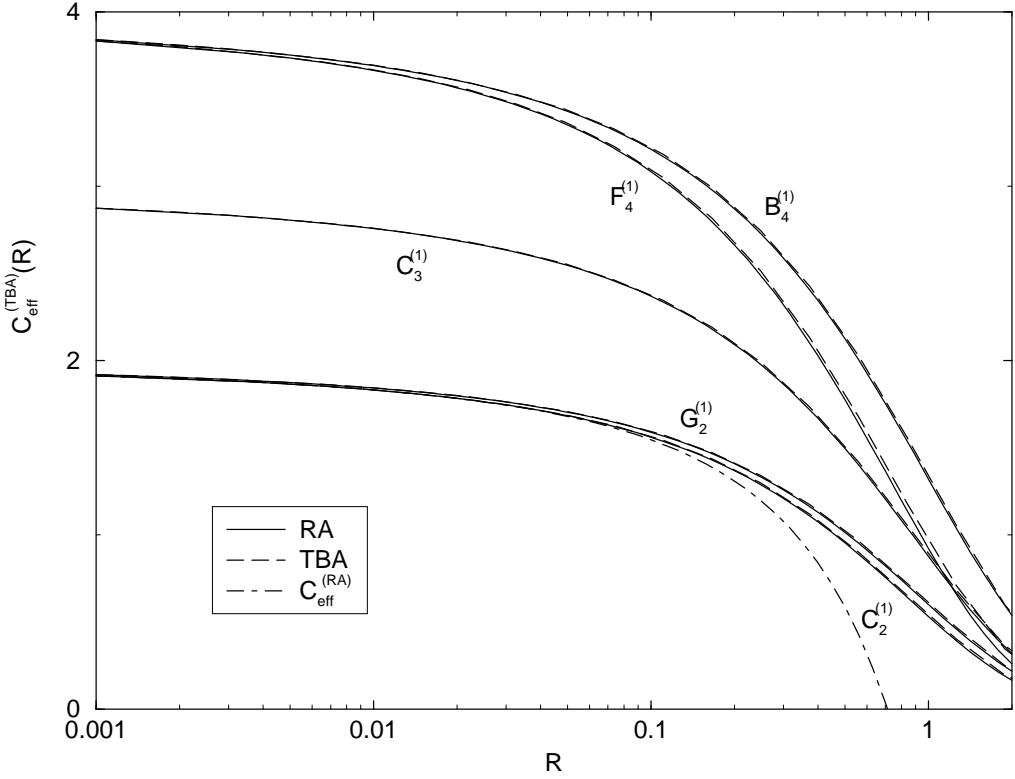}}}}}
\caption{Plot of $c_{\rm eff}^{\rm (TBA)}$ for $C_2^{(1)},\ C_3^{(1)},
B_4^{(1)}$, $G_2^{(1)}$ and $F_4^{(1)}$ ATFTs at $B=0.5$. 
(We omit in the figure $C_4^{(1)}$ and
$B_3^{(1)}$ cases not to make it too complicated.) As an
example, we also display $c_{\rm eff}^{\rm (RA)}$ for $C_2^{(1)}$  
calculated without taking into account the bulk term. The difference 
between this function and $c_{\rm eff}^{\rm (TBA)}$ gives the bulk free 
energy of $C_2^{(1)}$ ATFT according to Eq.(\ref{deltac}).}
\end{figure}

{\small\section{Concluding remarks}}

In the main part of this paper we considered the UV asymptotics
of the effective central charges in ATFTs. 
The most important CFT data, which we used
for this analysis were the reflection amplitudes (\ref{rhats}) of NATTs.
It was mentioned in Introduction, that these functions  play also a crucial
role in the calculation of the one point functions in perturbed CFT.
The one point functions of the exponential fields in ATFTs:
\beq \label{opf}
\T(\bds{a})= < \exp \bds{a}\cdot \bds{\varphi}>
\eeq
can be reconstructed from from the same reflection amplitudes. 
It follows from the results of the 
paper \cite{FLZZ} that functions (\ref{opf}) satisfy the functional
equations similar to the relations (\ref{reflection}) for the vertex operators.
These equations together with analyticity and symmetry conditions 
fix one point functions in perturbed CFTs. One can find the solution of these 
functional equations with proper analyticity properties and respecting 
all symmetries of extended Dynkin diagram of Lie algebra $G$. This solution 
is a natural generalization to the non-simply laced case of the one point 
function for $ADE$ series of ATFTs calculated in \cite{VF} and can be written
in the form:
\beaq \label{opfn} 
\T(\vaa) &=& \left[\frac{\mbar k(G)}{2} \Gamma\left(\frac{1-B}{H}\right)
\Gamma\left(1+\frac{B}{H}\right)\right]^{2\vQ\cdot\vaa-\vaa^2}
\prod_{i=1}^{r}[-\pi\mu_i\gamma(1+\bfe_i^2b^2/2)]^{-\vom^\vee_i\cdot\vaa/b}
\nonumber \\
&&\times 
   \exp \int \frac{dt}{t} 
      \left[ {\vaa}^2 e^{-2t} - {\cal F}({\vaa},t) \right] \,,
\eeaq 
where
\beq
{\cal F}(\vaa,t)=\sum_{{\val} >0}\frac{\sinh(a_{\val}bt)
\sinh((ba_{\val}-2bQ_{\val}+(1+b^2)H)t)\sinh((b^2\val^2/2+1)t)}
{\sinh{t}\sinh(b^2\val^2t/2)\sinh((1+b^2)Ht)}.
\eeq 
The one point function $\T(\vaa)$ can be used for the analysis of ATFTs. 
In particular, it contains the information about the bulk free energy
$f(G)$, which was calculated independently by Bethe Ansatz method. 
One can easily derive from Eqs.(\ref{action}) and (\ref{mmu}) that:
\beq
\frac{n_if(G)}{H(1+b^2)}=\mu_i\T(b\bfe_i)
\eeq
Using Eq.(\ref{opfn}) for function $\T(\vaa)$ one finds:
\beaq \label{int}
\frac{-4\pi\ga(1+\bfe_i^2b^2/2)n_if(G)}{H(1+b^2)(\mbar k(G))^2} &=&
\left[\Gamma\left(\frac{1-B}{H}\right)
\Gamma\left(1+\frac{B}{H}\right)\right]^2\nonumber\\
&&\times\exp \int \frac{dt}{t} 
      \left[ (b\bfe_i)^2 e^{-2t} - {\cal F}({b\bfe_i},t) \right]  
\eeaq
The integral in the exponent can be calculated and results coincides
with Eq.(\ref{fg}). This gives the nonperturbative test to the 
one point function $\T(\vaa)$. In particular, taking the limit
$b\rightarrow 0$ in Eq.(\ref{int}) (and the dual limit) 
one can derive the amusing relations
for gamma-functions associated with Lie algebras $G$. It is convenient 
to introduce the integers $n_i^{\vee}=n_i\bfe_i^2/2$ $i=0,\cdots,r$.  
Then these relations can be written as:
\beq
 \prod_{\val>0}\left(\ga(\val\cdot\vrh^\vee/h)\right)^{-\bfe_i\cdot\val^\vee}=
 n_i^{\vee}\left(\prod_{i=0}^{r}(n_i^{\vee})^{n_i}\right)^{-1/h},
\eeq
and
\beq
 \prod_{\val>0}\left(\ga(\val\cdot\vrh/h^\vee)\right)^{-\bfe_i^{\vee}\cdot\val}=
 n_i^{\vee}\bfe_i^2/2\left(\prod_{i=0}^{r}
(n_i^{\vee}\bfe_i^2/2)^{n_i^{\vee}}\right)^{-1/h^\vee}.
\eeq
More detailed consideration of one point functions in ATFTs 
we suppose to give in another publication. 

\section*{\bf Acknowledgement} 

We thank F. Smirnov and Al. Zamolodchikov for valuable discussions. 
PB gratefully thanks for the hospitality of KIAS where part of this work
was done. This work is supported in part by 
MOST 98-N6-01-01-A-05 (CA), Korea Research Foundation 
KRF-99-015-DI0021 (CA) 1998-015-D00071 (CR), 
and KOSEF 1999-2-112-001-5(CA,CR). 
PB's work is supported in part by the EU under contract ERBFMRX.
VF's work is supported in part by the EU under contract ERBFMRX CT960012.

\end{document}